\documentstyle[twoside,fleqn,espcrc2_pre,espcrc2,epsf]{article}

\newcommand{\vsaf}{\vspace*{-15pt}}

\begin{document}

\title{Feasibility study of using the overlap-Dirac operator
for hadron spectroscopy.\thanks{
Presented by C.\ McNeile.}
}

\author{UKQCD Collaboration, Craig~McNeile, 
Alan~Irving, and  Chris~Michael,\address{Department of Mathematical Sciences, 
University of\ Liverpool, L69 3BX, UK}
}

\begin{abstract}
 We investigate a number of algorithms that calculate the 
 quark propagators for the overlap-Dirac fermion operator.
  The QCD simulations were
  performed at $\beta = 5.9$ with a lattice volume of $16^3 \; 32$.

\end{abstract}
 
\maketitle

\section{INTRODUCTION}

Two critical systematic errors in the calculation of the $f_B$
decay constant from lattice QCD are the chiral extrapolations and the
unquenching errors~\cite{Bernard:1998xi}. The only way to reduce these
errors is to simulate QCD with lighter quark masses. Unfortunately,
because of exceptional configurations, it is difficult to
further reduce the masses of the light quarks in quenched simulations
with the clover operator. Progress in reducing the sea quark masses
in dynamical fermion simulations with Wilson like quarks is 
slow~\cite{Gottlieb:1997hy}.

It seems plausible that the difficulty of simulating with light quark
masses with the clover operator is due to explicit chiral symmetry
breaking in the action.  Neuberger has derived~\cite{Neuberger:1997fp}
a fermion operator, called the overlap-Dirac operator, that has a
lattice chiral symmetry~\cite{Ginsparg:1982bj,Luscher:1998pq}.

Our goal is to simulate the overlap-Dirac operator in a mass region:
($M_{PS}/M_V = 0.3 - 0.5$), which is inaccessible to clover quarks
(but not staggered quarks~\cite{Gottlieb:1997hy}). Most of the
techniques developed in the quenched theory can be used for full QCD
simulations~\cite{Kennedy:1998cu}.

\section{THE OVERLAP-DIRAC OPERATOR}

The massive overlap-Dirac
operator~\cite{Neuberger:1997fp,Edwards:1998wx} 
is 
\begin{equation}
D^{N} = \frac{1}{2}( 1 + \mu +
(1-\mu) \gamma_{5} \frac{H(m)}{\sqrt{H(m)H(m) }}     )
\end{equation}
where $H(m)$ is the hermitian Wilson fermion operator with negative
mass, defined by
\begin{equation}
H(m) = \gamma_5 ( D^W - m)
\label{eq:gfiveWilson}
\end{equation}
where $D^W$ is the standard Wilson fermion operator.
The parameter $\mu$ 
is related to the physical quark mass and lies in the range
$0$ to $1$.
The $m$ parameter is a regulating mass, in the range between a
critical value and 2.
The physics should be independent of the mass $m$, but the value of
$m$ effects the locality of the operator and the number of iterations
required in some of the algorithms used to compute the overlap-Dirac
operator.

\section{NUMERICAL TECHNIQUES}

Quark propagators are calculated using a sparse matrix inversion
algorithm.  The inner step of the inverter is the application of the
fermion matrix to a vector. For computations that use the 
overlap-Dirac,
the step function
\begin{equation}
\epsilon(H) \underline{b} = 
\frac{H} { \sqrt{H  H } }  \underline{b}
\label{eq:stepfunc}
\end{equation}
must be computed using some sparse matrix algorithm.
The nested nature of the algorithm required to calculate the 
quark propagators for the overlap-Dirac operator makes the 
simulations considerably more expensive than 
those that use traditional fermion operators.

Practical calculations of the overlap operator are necessarily approximate.  To
judge the accuracy of our approximate calculation we used the
(GW) Ginsparg-Wilson error:
\begin{equation}
\mid \mid 
 \gamma_5 D^N \underline{x} 
+   D^N  \gamma_5  \underline{x} 
   - 
  2 D^N  \gamma_5 D^N  \underline{x} 
\mid \mid 
\frac{1}{\mid\mid \underline{x} \mid\mid  }
\label{eq:GWerrDEFN}
\end{equation}
which just checks that the matrix obeys the 
Ginsparg-Wilson relation~\cite{Ginsparg:1982bj}.


Our numerical simulations were done using $\beta = 5.9$ quenched 
gauge configurations, with a volume of $16^3 32$. The quark
propagators were generated from point sources. For all the algorithms
we investigated, we used $m$ equal to $1.5$.

\section{LANCZOS BASED METHOD}

Borici~\cite{Borici:1998mr} has developed a method 
to calculate the action of the overlap-Dirac operator on a vector, using
the Lanczos algorithm.
In exact arithmetic,
the Lanczos algorithm generates an orthonormal
set of vectors that tridiagonalises the matrix.
\begin{equation}
H Q_n = Q_n T_n 
\end{equation}
where $T_n$ is a tridiagonal matrix. The columns of $Q_n$ contain
the Lanczos vectors.

The ``trick'', to evaluate the step function (Eq.~\ref{eq:stepfunc}),
is to set the target 
vector $\underline{b}$, as the first vector in the Lanczos 
sequence.
An arbitrary function $f$ of the matrix $H$ 
acting on a vector is constructed using
\begin{eqnarray}
(f(H) b )_i & = & \sum_{j} ( Q_n f(T_n) Q_n^\dagger )_{i\; j} b_j \\
& = & \| b \|  ( Q_n f(T_n)  )_{i\; 1} \label{eq:what_to_code}
\end{eqnarray}
where the orthogonality of the Lanczos vectors has been used.
The
$f(T_n)$ matrix is computed using standard dense linear algebra
routines. For the step function the eigenvalues of $T_n$ are replaced
by their moduli. Eq.~\ref{eq:what_to_code} is linear
in the Lanczos vectors and thus can be computed in two passes.

The major problem with the Lanczos procedure is the loss of the
orthogonality of the sequence of vectors due to rounding errors. It is
not clear how this lack of orthogonality effects the final results.
Some theoretical analysis has been done on this
method~\cite{Druskin:1998:UNL}. It is claimed that the lack of
orthogonality is not important for some classes of functions.

\begin{figure}
\vbox{\epsfxsize=2.7in \epsfbox{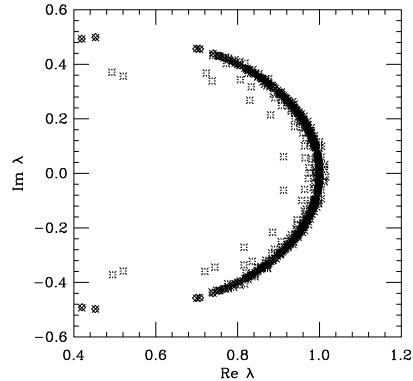}}
\caption{Eigenvalue spectrum of the overlap-Dirac operator,
for 100 (crosses), 50 (diamonds), and 10 (bursts) Lanczos
iterations.}
\label{fig:circle}
\vskip -0.5in
\vsaf
\end{figure}

In Fig.~\ref{fig:circle}, we plot the eigenvalues of the
Ginsparg-Wilson operator, as a function of the number of Lanczos
steps, for a $2^4$ hot $SU(3)$ configuration.  As the number of
Lanczos steps increases, the eigenvalue spectrum moves closer to
a circle (the correct result). Even after 50 iterations of the Lanczos
algorithm, there are still small deviations from the circle.
Unfortunately, it is much harder to look at the eigenvalue spectrum
using a production gauge configuration, so we computed the  GW error 
instead. The GW error was: $5\; 10^{-3}$ (50 iterations), $6\; 10^{-4}$ (100)
iterations, and $3\; 10^{-4}$ (300 iterations) on a single gauge
configuration.

\begin{figure}
\vbox{\epsfxsize=2.7in \epsfbox{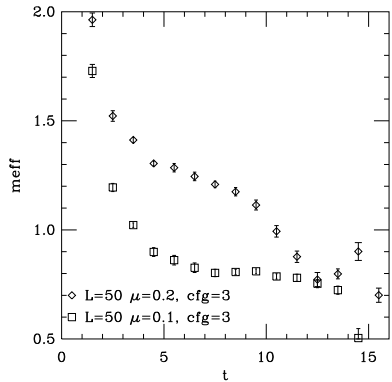}}
\vskip -9mm
\caption{Pion effective mass plot}
\label{fig:meff}
\vskip -8mm
\end{figure}

Fig.~\ref{fig:meff}, is an effective mass plot of the pion, for two
choices of mass and number of Lanczos sweeps.  The ``plateau'' in
the pion effective mass plot for approximate 
operator that used 50 Lanczos iterations is
higher than the lowest pion mass that can be reached with
non-perturbatively improved clover.  It is not clear what causes the
"shoulder" in Fig.~\ref{fig:meff}. We would like to compute the
eigenvalues of the overlap-Dirac operator on the bigger gauge configuration,
to check how accurately we are computing the overlap-Dirac operator.

\section{RATIONAL APPROXIMATION}

The step function can be be approximated by a rational
approximation.
\begin{equation}
\epsilon(H) \sim H \; ( c_0 + \sum_{k=1}^{N} 
   \frac{c_k}{H^2 + d_k}  )
\label{eq:pole}
\end{equation}
The rational approximation typically approximates the step function,
between two values. The eigenvalues of the matrix $H$ should lie in
the region where the approximation is good. The coefficients $c_k$ and
$d_k$ can be obtained from the Remez algorithm~\cite{Edwards:1998yw}.
The number of iterations required in the inverter is controlled by the
smallest $d_k$ coefficient, which acts like a mass. We have not yet
implemented the technique of projecting out some of the low lying
eigenmodes~\cite{Edwards:1998yw}.

On one configuration we obtained GW errors of: $1\; 10^{-4}$, and $4 \;
10^{-5}$, for the $N=6$, and $N=8$, optimal
rational approximations~\cite{Edwards:1998yw}.  The multiplicative
scaling factor was tuned to obtain the best results.  Unfortunately,
the above results required up to 600 iterations for the 
smallest $d_k$, which was too large
to use as the inner step of a quark propagator inverter.

One feature of the optimal rational
approximation~\cite{Edwards:1998yw}, is that the lowest $d_k$ factor
is smaller than the square of the validity of the approximation, which
means that the condition number of the inversion is that of the matrix
$H^2$. We experimented with a hybrid quadrature and series
approximation to Robertson's integral representation of the step
function.
\begin{eqnarray}
\epsilon(H) \!\!\! & = & \!\!\! \int_{0}^{\infty} 
\frac{2 H}{ \pi ( t^2 + H^2)}   dt \label{eq:hybrid} \\
            \!\!\! & \sim & \!\!\!
\int_{0}^{\theta_S} 
\frac{2 H}{ \pi ( t^2 + H^2)}   dt  
+
\int_{\theta_S}^{\theta_L} 
\frac{2 H}{ \pi ( t^2 + H^2)}   dt  
\nonumber
\end{eqnarray}
In Eq.~\ref{eq:hybrid}, the first integral was approximated using an
open quadrature rule and the second integral was approximated by a
Chebyshev series. The step size in the quadrature formulae reduces
the condition number of the required inversion. In our preliminary
tests of the algorithm, the hybrid method produced a substantial
reduction in the number of iterations required over the optimal
rational approximation. However the computed solution was less
accurate than that produced by the optimal rational approximation,
because the Ginsparg-Wilson error was $6 \; 10^{-4}$.

Clearly more work is required on the algorithms that
calculate the step function, before the overlap-Dirac
operator can be used in the quark mass region we are 
interested in.

%
%

This work is supported by PPARC.
The computations were carried out on the T3D and T3E at EPCC
in Edinburgh.



\end{document}